\documentclass[twocolumn]{aastex63}
\usepackage[utf8]{inputenc}

\usepackage{natbib}
\usepackage{graphicx}
\usepackage{amsmath}
\usepackage{booktabs}
\usepackage{xspace}
\usepackage{hyperref}
\usepackage[T1]{fontenc}
\usepackage{lipsum}
\usepackage{comment}
\usepackage{xcolor}
\usepackage{multirow}
\usepackage{enumitem}

\usepackage{IEEEtrantools}
\usepackage{amsmath}

\providecommand{\mj}{\ensuremath{\,M_{\rm J}}}

\newcommand{\rearth}{\mbox{$R_{\ensuremath{\oplus}}$}}
\newcommand{\mearth}{\mbox{$M_{\ensuremath{\oplus}}$}}

\newcommand{\rBb}{\ensuremath{1.29 \pm 0.55}}
\newcommand{\pBb}{\ensuremath{3.09307 \pm 0.00024}}
\newcommand{\pBc}{\ensuremath{2100.6 \pm 2.9}}
\newcommand{\mBc}{\ensuremath{0.308 \pm 0.014}}

\newcommand{\rCb}{\ensuremath{2.40 \pm 0.18}}
\newcommand{\pCb}{\ensuremath{2.68005 \pm 0.00003}}
\newcommand{\mCb}{\ensuremath{5.2 \pm 3.1}}
\newcommand{\rhoCb}{\ensuremath{2.06 \pm 1.13}}
\newcommand{\pCc}{\ensuremath{502 \pm 16}}
\newcommand{\mCc}{\ensuremath{0.573 \pm 0.074}}
\newcommand{\eCc}{\ensuremath{0.14 \pm 0.13}}

\newcommand{\gdotC}{\ensuremath{-0.0286 \, \pm \, 0.0047}}
\newcommand{\gdotCecc}{\ensuremath{-0.0261 \, \pm \, 0.0058}}
\newcommand{\NobsC}{\ensuremath{20}}

\newcommand{\rDb}{\ensuremath{5.44 \pm 0.18}}
\newcommand{\pDb}{\ensuremath{3.77015 \pm 0.00010}}
\newcommand{\mDb}{\ensuremath{26.1 \pm 2.2}}
\newcommand{\rhoDb}{\ensuremath{0.89 \pm 0.12}}
\newcommand{\pDc}{\ensuremath{389.2  \pm 3.9}}
\newcommand{\mDc}{\ensuremath{1.05 \pm 0.05}}
\newcommand{\eDc}{\ensuremath{0.18 \pm 0.05}}

\newcommand{\NobsD}{\ensuremath{20}}

\newcommand{\rEb}{\ensuremath{2.02 \pm 0.26}}
\newcommand{\pEb}{\ensuremath{0.73649 \pm 0.00002}}
\newcommand{\pEc}{\ensuremath{5285 \pm 5}}
\newcommand{\mEc}{\ensuremath{3.84 \pm 0.08}}

\begin{document}

\title{TESS-Keck Survey XIV: Two giant exoplanets from the Distant Giants Survey}

\author[0000-0002-4290-6826]{Judah Van Zandt}
\affiliation{Department of Physics \& Astronomy, University of California Los Angeles, Los Angeles, CA 90095, USA}

\author[0000-0003-0967-2893]{Erik A. Petigura}
\affiliation{Department of Physics \& Astronomy, University of California Los Angeles, Los Angeles, CA 90095, USA}

\author[0000-0003-2562-9043]{Mason MacDougall}
\affiliation{Department of Physics \& Astronomy, University of California Los Angeles, Los Angeles, CA 90095, USA}

\author[0000-0003-0742-1660]{Gregory J. Gilbert}
\affiliation{Department of Physics \& Astronomy, University of California Los Angeles, Los Angeles, CA 90095, USA}

\author[0000-0001-8342-7736]{Jack Lubin}
\affiliation{Department of Physics \& Astronomy, University of California Irvine, Irvine, CA 92697, USA}

\author[0000-0001-7139-2724]{Thomas~Barclay}
\affiliation{NASA Goddard Space Flight Center, 8800 Greenbelt Road, Greenbelt, MD 20771, USA}
\affiliation{University of Maryland, Baltimore County, 1000 Hilltop Circle, Baltimore, MD 21250, USA}

\author[0000-0002-7030-9519]{Natalie M. Batalha}
\affiliation{Department of Astronomy and Astrophysics, University of California, Santa Cruz, CA 95064, USA}

\author{Ian J. M. Crossfield}
\affiliation{Department of Physics \& Astronomy, University of Kansas, 1082 Malott, 1251 Wescoe Hall Dr., Lawrence, KS 66045, USA}

\author[0000-0001-8189-0233]{Courtney Dressing}
\affiliation{Department of Astronomy, 501 Campbell Hall, University of California, Berkeley, CA 94720, USA}

\author[0000-0003-3504-5316]{Benjamin Fulton}
\affiliation{NASA Exoplanet Science Institute/Caltech-IPAC, MC 314-6, 1200 E. California Blvd., Pasadena, CA 91125, USA}

\author[0000-0001-8638-0320]{Andrew W. Howard}
\affiliation{Department of Astronomy, California Institute of Technology, Pasadena, CA 91125, USA}

\author[0000-0001-8832-4488]{Daniel Huber}
\affiliation{Institute for Astronomy, University of Hawai`i, 2680 Woodlawn Drive, Honolulu, HI 96822, USA}

\author[0000-0002-0531-1073]{Howard Isaacson}
\affiliation{Department of Astronomy, 501 Campbell Hall, University of California, Berkeley, CA 94720, USA}
\affiliation{Centre for Astrophysics, University of Southern Queensland, Toowoomba, QLD, Australia}

\author[0000-0002-7084-0529]{Stephen R. Kane}
\affiliation{Department of Earth and Planetary Sciences, University of California, Riverside, CA 92521, USA}

\author[0000-0003-0149-9678]{Paul Robertson}
\affiliation{Department of Physics \& Astronomy, University of California Irvine, Irvine, CA 92697, USA}

\author[0000-0001-8127-5775]{Arpita Roy}
\affiliation{Space Telescope Science Institute, 3700 San Martin Drive, Baltimore, MD 21218, USA}
\affiliation{Department of Physics and Astronomy, Johns Hopkins University, 3400 N Charles St, Baltimore, MD 21218, USA}

\author[0000-0002-3725-3058]{Lauren M. Weiss}
\affiliation{Department of Physics and Astronomy, University of Notre Dame, Notre Dame, IN 46556, USA}

\author[0000-0003-0012-9093]{Aida Behmard}
\altaffiliation{NSF Graduate Research Fellow}
\affiliation{Division of Geological and Planetary Science, California Institute of Technology, Pasadena, CA 91125, USA}

\author[0000-0001-7708-2364]{Corey Beard}
\affiliation{Department of Physics \& Astronomy, University of California Irvine, Irvine, CA 92697, USA}

\author[0000-0003-1125-2564]{Ashley Chontos}
\altaffiliation{Henry Norris Russell Postdoctoral Fellow}
\affiliation{Department of Astrophysical Sciences, Princeton University, 4 Ivy Lane, Princeton, NJ 08540, USA}
\affiliation{Institute for Astronomy, University of Hawai`i, 2680 Woodlawn Drive, Honolulu, HI 96822, USA}

\author[0000-0002-8958-0683]{Fei Dai}
\altaffiliation{NASA Sagan Fellow}
\affiliation{Division of Geological and Planetary Sciences,
1200 E California Blvd, Pasadena, CA, 91125, USA}
\affiliation{Department of Astronomy, California Institute of Technology, Pasadena, CA 91125, USA}

\author[0000-0002-4297-5506]{Paul A.\ Dalba}
\altaffiliation{Heising-Simons 51 Pegasi b Postdoctoral Fellow}
\affiliation{Department of Astronomy and Astrophysics, University of California, Santa Cruz, CA 95064, USA}
\affiliation{SETI Institute, Carl Sagan Center, 339N Bernardo Avenue, Suite 200, Mountain View, CA 94043, USA}

\author[0000-0002-3551-279X]{Tara Fetherolf}
\altaffiliation{UC Chancellor's Fellow}
\affiliation{Department of Earth and Planetary Sciences, University of California, Riverside, CA 92521, USA}

\author[0000-0002-8965-3969]{Steven Giacalone}
\affiliation{Department of Astronomy, 501 Campbell Hall, University of California, Berkeley, CA 94720, USA}

\author{Christopher~E.~Henze}
\affiliation{NASA Ames Research Center, Moffett Field, CA 94035, USA}

\author[0000-0002-0139-4756]{Michelle L. Hill}
\affiliation{Department of Earth and Planetary Sciences, University of California, Riverside, CA 92521, USA}

\author{Lea A.\ Hirsch}
\affiliation{Kavli Institute for Particle Astrophysics and Cosmology, Stanford University, Stanford, CA 94305, USA}

\author[0000-0002-5034-9476]{Rae Holcomb}
\affiliation{Department of Physics \& Astronomy, University of California Irvine, Irvine, CA 92697, USA}

\author[0000-0002-2532-2853]{Steve~B.~Howell}
\affil{NASA Ames Research Center, Moffett Field, CA 94035, USA}

\author[0000-0002-4715-9460]{Jon M. Jenkins}
\affiliation{NASA Ames Research Center, Moffett Field, CA 94035, USA}

\author[0000-0001-9911-7388]{David W. Latham}
\affiliation{Center for Astrophysics | Harvard \& Smithsonian, 60 Garden St, Cambridge, MA 02138, USA}

\author{Andrew Mayo}
\affiliation{Department of Astronomy, 501 Campbell Hall, University of California, Berkeley, CA 94720, USA}

\author[0000-0002-4510-2268]{Ismael~Mireles}
\affiliation{Department of Physics and Astronomy, University of New Mexico, 210 Yale Blvd NE, Albuquerque, NM 87106, USA}

\author[0000-0003-4603-556X]{Teo Mo\v{c}nik}
\affiliation{Gemini Observatory/NSF's NOIRLab, 670 N. A'ohoku Place, Hilo, HI 96720, USA}

\author[0000-0001-8898-8284]{Joseph M. Akana Murphy}
\altaffiliation{NSF Graduate Research Fellow}
\affiliation{Department of Astronomy and Astrophysics, University of California, Santa Cruz, CA 95064, USA}

\author[0000-0001-9771-7953]{Daria Pidhorodetska}
\affiliation{Department of Earth and Planetary Sciences, University of California, Riverside, CA 92521, USA}

\author[0000-0001-7047-8681]{Alex S. Polanski}
\affiliation{Department of Physics \& Astronomy, University of Kansas, 1082 Malott, 1251 Wescoe Hall Dr., Lawrence, KS 66045, USA}

\author[0000-0003-2058-6662]{George R. Ricker}
\affiliation{Department of Physics and Kavli Institute for Astrophysics and Space Research, Massachusetts Institute of Technology, Cambridge, MA 02139, USA}

\author{Lee J.\ Rosenthal}
\affiliation{Department of Astronomy, California Institute of Technology, Pasadena, CA 91125, USA}

\author[0000-0003-3856-3143]{Ryan A. Rubenzahl}
\altaffiliation{NSF Graduate Research Fellow}
\affiliation{Department of Astronomy, California Institute of Technology, Pasadena, CA 91125, USA}

\author[0000-0002-6892-6948]{S. Seager}
\affil{Department of Earth, Atmospheric, and Planetary Sciences, Massachusetts Institute of Technology, Cambridge, MA 02139, USA}
\affil{Department of Physics and Kavli Institute for Astrophysics and Space Research, Massachusetts Institute of Technology, Cambridge, MA 02139, USA}
\affil{Department of Aeronautics and Astronautics, Massachusetts Institute of Technology, Cambridge, MA 02139, USA}

\author[0000-0003-3623-7280]{Nicholas Scarsdale}
\affiliation{Department of Astronomy and Astrophysics, University of California, Santa Cruz, CA 95064, USA}

\author[0000-0002-1845-2617]{Emma V. Turtelboom}
\affiliation{Department of Astronomy, 501 Campbell Hall, University of California, Berkeley, CA 94720, USA}

\author[0000-0001-6763-6562]{Roland Vanderspek}
\affiliation{Department of Physics and Kavli Institute for Astrophysics and Space Research, Massachusetts Institute of Technology, Cambridge, MA 02139, USA}

\author[0000-0002-4265-047X]{Joshua N. Winn}
\affiliation{Department of Astrophysical Sciences, Princeton University, 4 Ivy Lane, Princeton, NJ 08544, USA}



\begin{abstract}
We present the Distant Giants Survey, a three-year radial velocity (RV) campaign to measure P(DG|CS), the conditional occurrence of distant giant planets (DG; $M_p \sim 0.3 - 13 \, \mj$, $P > 1 \, \text{year}$) in systems hosting a close-in small planet (CS; $R_p < 10 \, \rearth$). For the past two years, we have monitored 47 Sun-like stars hosting small transiting planets detected by TESS. We present the selection criteria used to assemble our sample and report the discovery of two distant giant planets, TOI-1669 b and TOI-1694 c. For TOI-1669 b we find that $M\sin i = \mCc \, \mj$,  $P = \pCc$ days, and $e < 0.27$, while for TOI-1694 c, $M\sin i = \mDc \, \mj$, $P = \pDc$ days, and $e = \eDc$. We also confirmed the 3.8-day transiting planet TOI-1694 b by measuring a true mass of $M = \mDb \, \mearth$. At the end of the Distant Giants Survey, we will incorporate TOI-1669 b and TOI-1694 c into our calculation of P(DG|CS), a crucial statistic for understanding the relationship between outer giants and small inner companions.

\end{abstract}


\section{Introduction}
The past 30 years of exoplanet discovery have revealed a variety of distinct planet classes. The most abundant of these discovered to date around Sun-like stars are between the size of Earth and Neptune with orbital periods of a year or less. Statistical analyses of Kepler data \citep{Borucki2010} have shown that such planets occur at a rate of $\sim1$ per star (see, e.g., \cite{Petigura2018}). Meanwhile, ground-based RV surveys (e.g., \citealt{Rosenthal2021b, Cumming2008, Fischer2014, Wittenmyer2016}) report that long-period ($P \gtrsim 1 \, \text{year}$) giant planets are somewhat rare, orbiting $\sim$5-20\% of Sun-like stars. However, the distinct observing strategies employed by Kepler versus RV surveys produced stellar samples with little overlap. On the one hand, Kepler continuously monitored $>10^5$ stars along a fixed line of sight; the typical planet host in this sample is 600 parsecs from Earth with a brightness of $V=14$. By contrast, ground-based RV surveys have targeted bright, nearby stars that are distributed roughly evenly on the sky; the typical planet host in this sample is 40 parsecs from Earth with a brightness of $V=8$. Because the inner transiting planets mostly discovered by Kepler and the outer giants mostly discovered by RVs are drawn from nearly disjoint stellar samples, the connection between them is unclear.

Current planet formation models differ on whether the processes that produce long-period gas giants and close-in small planets are positively or negatively correlated. Strict in-situ models (e.g., \citealt{Chiang2013}) predict that the metal-rich protoplanetary disks known to facilitate gas giant formation \citep{FischerValenti2005, Mordasini2009} also promote the growth of sub-Jovian cores at close separations. On the other hand, models involving significant planetary migration predict an \textit{anti}-correlation, where nascent planetary cores either 1) develop beyond the ice line and are blocked from inward migration by newly-formed giants at a few AU \citep{Izidoro2015}, or 2) develop close in and are driven into their host star by inward giant planet migration \citep{Batygin2015}.

Recent observational works have directly estimated the conditional occurrence of distant giant companions to close-in small planets, P(DG|CS). \cite{Zhu_DG_2018} and \cite{Bryan2019}, hereafter Z18 and B19, respectively, each analyzed archival RV data for systems with super-Earths. Z18 estimated P(DG|CS)$\approx30\%$ using the following procedure: first, they counted the known systems with a Sun-like host, at least one inner super-Earth ($R_{p} < 4 \, \rearth$, P$<400$ days) and an RV baseline $>1$ year; then they divided the number of these systems reported to host a distant giant by the total. B19 estimated P(DG|CS)$=39\pm7\%$ using a similar procedure. They selected systems with at least one confirmed super-Earth and at least 10 RVs over 100 days. Unlike Z18, they re-fit those RVs using \texttt{radvel} \citep{radvel} to search for unknown companions, considering both full and partial orbits. Both analyses indicate a factor of $\sim$3--4 enhancement over the field occurrence rate, but are vulnerable to systematic biases due to their loosely-defined target selection functions and heterogeneity of RV time series (quality, sampling strategy, and baseline). In particular, Z18 and B19 selected targets where significant RV baselines had already been collected by other surveys. However, earlier studies may have chosen their RV targets based on a variety of criteria, including an increased probability of hosting planets. The aggregation of RV targets from separate studies may bias the associated planet populations, and because Z18 and B19 did not address these factors on a target-by-target basis, the extent to which this bias may have influenced their final results is unclear.

A more uniform analysis was carried out as part of the California Legacy Survey (CLS; \citealt{Rosenthal2021b}). The CLS sample consists of 719 Sun-like stars with similar RV baselines and precisions, and chosen without bias toward stars with a higher or lower probability of hosting planets. Furthermore, \cite{Rosenthal2021b} performed a uniform iterative search for periodic signals in each RV time series using the \texttt{rvsearch} package \citep{Rosenthal2021a}, recovering populations of both inner small planets ($0.023-1 $ AU, $2-30 \, \mearth$) and outer giants ($0.23-10$ AU, $30-6000 \, \mearth$). The authors measured a conditional occurrence of P(DG|CS)=41$\pm$15\%. Although this value is consistent with the findings of both Z18 and B19, \cite{Rosenthal2021b} also found a prior distant giant occurrence of P(DG)=17.6$\pm$2.2\%, meaning that their conditional occurrence is $\sim$1.6$\sigma$ separated from a null result.

We present the Distant Giants Survey, a 3-year RV survey to determine P(DG|CS) in a sample of Sun-like transiting planet hosts from the TESS mission \citep{Ricker2015}.\footnote{Because our survey requires that all systems host a transiting inner planet, we are actually constraining the conditional occurrence of distant giant companions to \textit{transiting} close-in small ones. However, we expect the population of stars with transiting inner small planets to host outer giants at the same rate as stars hosting inner planets irrespective of a transiting geometry. On the other hand, if systems hosting both planet types tend to be coplanar, we will have greater RV sensitivity to giants in transiting systems. We will account for the resulting bias in detail in our statistical analysis.} In designing our survey, we took care to construct a uniform stellar sample to avoid bias against or in favor of stars that host outer giant planets. We also applied a single observing strategy to achieve uniform planet sensitivity across our sample. Since beginning the survey in mid 2020, we have found evidence for 11 outer companions, both as resolved (i.e., complete) orbits and long-term trends. Distant Giants is part of the larger TESS-Keck Survey (TKS; \citealt{Chontos2021}), a multi-institutional collaboration to explore exoplanet compositions, occurrence, and system architectures (see, e.g., \citealt{Dalba2020}, \citealt{Weiss2021}, \citealt{Rubenzahl2021}).

In this paper, we introduce the Distant Giants Survey and highlight two new giant planets, TOI-1669 b and TOI-1694 c, detected in our sample. In Section \ref{sec:survey_design}, we describe the Distant Giants Survey as a whole, including our target selection process, observing strategy, and procedure for obtaining precise RVs from Keck-HIRES. Sections \ref{sec:TOI-1669} and \ref{sec:TOI-1694} detail our analysis of TOI-1669 and TOI-1694, including our RV model and the properties of the planets in each system. In Section \ref{sec:discussion}, we discuss our findings. In Section \ref{sec:conclusion}, we summarize our results and outline future work.


\section{Distant Giants Survey Design}
\label{sec:survey_design}

\subsection{Target Selection}
\label{subsec:target_selection}
Our ability to draw robust statistical conclusions from the Distant Giants Survey relies critically on the assembly of a well-defined stellar sample. We designed our target selection criteria to yield a sample of Sun-like stars hosting at least one small transiting planet found by TESS. We also required that all targets be amenable to precise RV follow-up from the northern hemisphere. To impose these criteria, we began with the master target list produced by \cite{Chontos2021}, which contains 2136 individual TESS Objects of Interest (TOIs) among 2045 planetary systems. We then applied the following sets of filters:

\begin{enumerate}
    \item \textit{Photometric and astrometric measurements.} To allow for efficient observation from the Northern Hemisphere, we required that all stars have $\delta > 0^{\circ}$ and $V<12.5$, where $\delta$ is the declination and $V$ is the V-band magnitude. We excluded stars with a \textit{Gaia} Renormalized Unit Weight Error (RUWE; \citealt{Lindegren2018}, \citealt{GaiaEDR3}) greater than 1.3 to ensure precise fits to Gaia's 5-parameter astrometric model. $\text{RUWE}<1.2$ is a conservative limit to exclude binary systems mis-classified as single sources (\citealt{Bryson2020}, Kraus et al. in preparation). All but one target in the Distant Giants sample satisfy $\text{RUWE}<1.2$, implying a low probability that we included any unwanted binary systems. We chose an upper bound of $10 \, R_{\Earth}$ on the transiting planet radius. In the event that a star hosted multiple transiting planets, we required that at least one meet our planet size requirement. These cuts reduced the target pool from 2045 to 147 systems.
    
    \item \textit{Data Quality.} We evaluated transit quality using TESS data validation (DV) reports retrieved from the TOI Catalog in July 2020. To ensure high-significance transits, we only included TOIs with at least one light curve produced by the TESS Science Processing Operations Center pipeline (SPOC, \citealt{Jenkins2016}). Furthermore, we used the Multiple Event Statistic (MES, \citealt{Jenkins2002}), a proxy for the signal-to-noise ratio (SNR), to evaluate transit quality. After visual inspection of a subset of SPOC transit fits, we found that $\text{MES} = 12$ was a suitable lower limit for identifying compelling detections. We also excluded targets with close visual companions, which we defined as companions within 4" and 5 V-band magnitudes. These filters further reduced the pool from 147 to 67 systems.
    
    \item \textit{Inactive and slowly-rotating stars.} We used \texttt{SpecMatch} \citep{Petigura2017} to analyze each target's "template" spectrum (\S \ref{subsec:observations}). \texttt{SpecMatch} interpolates over a grid of synthetic stellar spectra to estimate stellar parameters such as effective temperature $\text{T}_{\text{eff}}$, projected rotational velocity $v\sin(i)$, surface gravity $\log g$, and metallicity [Fe/H]. For stars cooler than 4700 K, we used \texttt{SpecMatch-Emp} \citep{Yee2017} to interpolate over real spectra of K and M dwarfs, which are more reliable than model spectra at low temperatures. We excluded rapidly rotating stars ($v\sin(i) > 5$ km/s), as well as those with $\text{T}_{\text{eff}}$ above the Kraft break ($\sim6250$ K); such stars offer limited RV precision due to Doppler broadening \citep{Kraft1967}. We also derived stellar mass according to the methods of \cite{FulPet2018}, and selected only main sequence stars between $0.5$ and $1.5$  $M_{\odot}$, consistent with our solar analog requirement.
    
    We measured each star's chromospheric activity through its $\log R'_{\text{HK}}$ index \citep{Isaacson2010, Noyes1984}. This value quantifies the emission in the cores of the Calcium II H and K lines relative to the total bolometric emission of the star, with higher core emission corresponding to enhanced activity and therefore greater RV variability in the epoch that the activity is measured. We required that $\text{log}R'_{\text{HK}} < -4.7$. This limit, adopted from \cite{Howard2010c}, restricts our sample to `inactive' and `very inactive' stars, as defined by \cite{Henry1996}. We note that restricting $\log R'_{\text{HK}}$ and $v\sin(i)$ introduces a bias toward older stars due to the correlations between age and both Calcium H and K line emission and rotation speed (e.g., \citealt{Soderblom1991, Noyes1984}). We retained these filters to ensure RV quality, and will account for the associated bias in our final results.
    
    We applied this $\log R'_{\text{HK}}$ filter using available activity values in mid-2020, but because stellar activity is variable, some of our targets fluctuate above the $\text{log}R'_{\text{HK}} = -4.7$ limit. Furthermore, two targets, TOI-1775 and TOI-2088, did not have available activity values at the time we applied this filter. They were thus not excluded based on activity, and we retained them in the sample. Since then, we have found $\text{log}R'_{\text{HK}} = -4.72$ for TOI-2088 and $-4.28$ for TOI-1775. Knowing that TOI-1775 fails our $\text{log}R'_{\text{HK}}$ cut, we will carefully monitor its activity against any signals that develop in the RVs. Of the remaining 67 systems, 48 passed these filters. Due to time constraints, 47 of these were selected for TKS by the target prioritization algorithm detailed in \cite{Chontos2021}. The filters of both Distant Giants and TKS are given in Table \ref{tab:cuts}.
\end{enumerate}

\begin{figure}[t]
  \centering\includegraphics[width=0.50\textwidth]{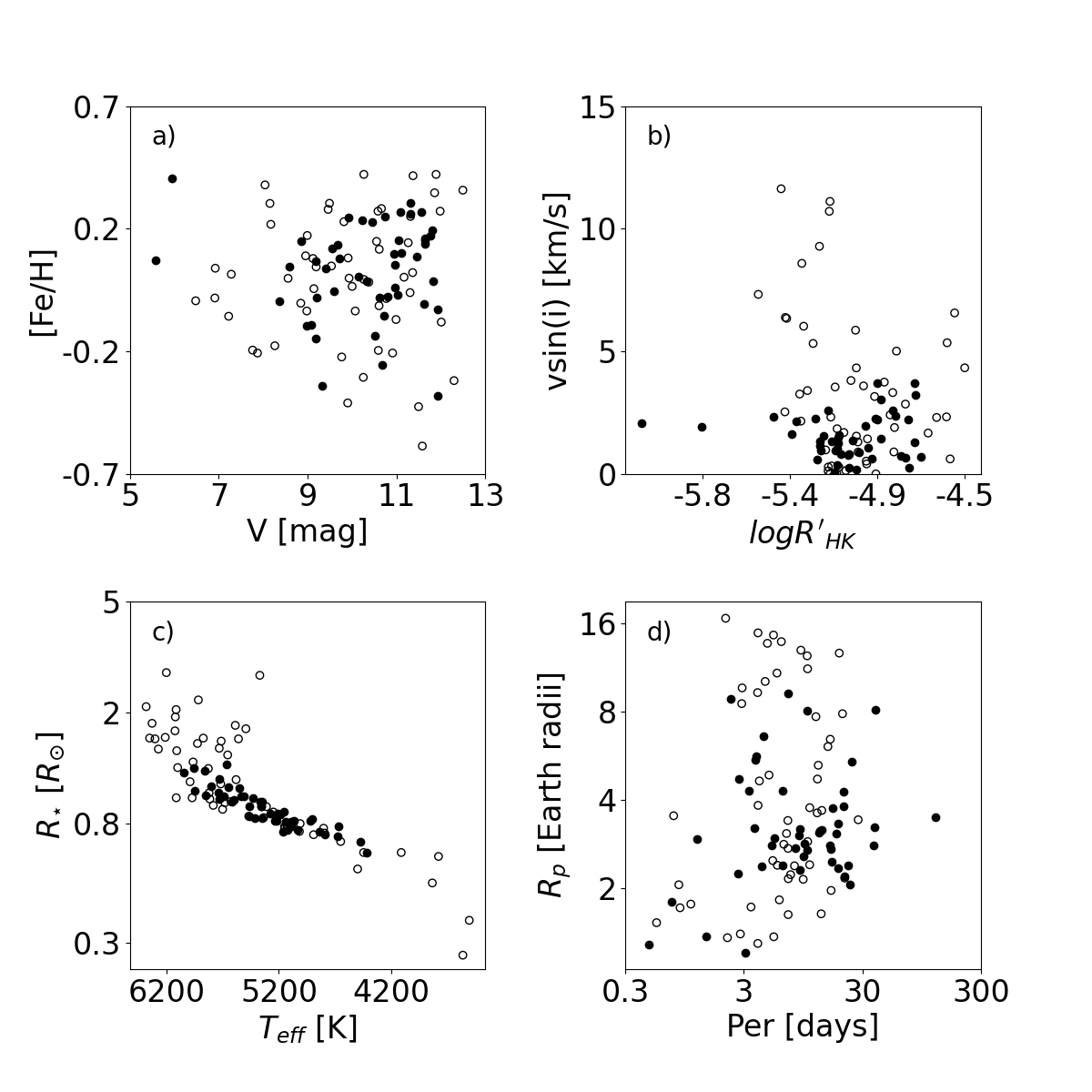}
 \centering\caption{Stellar and transiting planet parameters of the TKS survey. Filled points show targets selected for the Distant Giants sample. Panels a) through c) show stellar parameters. TOI-1775 and TOI-2088 are not shown in panel b) because they lacked measured $\text{log}R'_{\text{HK}}$ values when we finalized our sample. Panel d) shows parameters of the transiting planets. For multi-transiting Distant Giants systems, we checked planet radii in order of ascending TOI number, and show the first planet to pass our survey filters.}
  \label{fig:DG_multiplot}
\end{figure}

From the original set of 2045 TOI-hosting systems, 86 were ultimately selected for TKS and 47 were selected for Distant Giants. The final Distant Giants sample is given in Table \ref{tab:dg_transit_table}, and we summarize key stellar and planetary parameters of both Distant Giants and TKS in Figure \ref{fig:DG_multiplot}.

\begin{deluxetable*}{lcccc}
\tablewidth{0pt}
\caption{\textsc{Survey Criteria}}
\label{tab:cuts}

\tablehead{
\colhead{} & \colhead{} & \multicolumn{3}{c}{Distant Giants Survey}\vspace{-0.4cm} \\
\colhead{Parameter} & \colhead{TKS} & \multicolumn{3}{c}{} \vspace{-0.2cm}\\
\cline{3-5}
\colhead{} & \colhead{} & \colhead{Photometric} & \colhead{Manual} & \colhead{Spectroscopic}}
\startdata
Declination    & > $-30^{\circ}$    & > $0^{\circ}$ & --- & ---\\
$V$ & < 13.0                & < 12.5     & --- & ---\\
Evolutionary State  & MS or SG                    & MS               & --- & ---\\
RUWE              & < 2                & < 1.3      & --- & ---\\
$R_{\text{P}}$      & < 22 R$_{\Earth}$  & < 10 R$_{\Earth}$ & --- & ---\\
Transit Pipeline  & ---                          & --- & SPOC & ---\\
Detection Significance & SNR > 10   & --- & MES > 12 & ---\\
Close Companion   & $\Delta$V > 5 or sep > 2" & --- & $\Delta$V > 5 or sep > 4" & ---\\
$M_{\star}$    & ---                             & --- & --- & 0.5 $M_{\Sun}$ < $M_{\star}$ < 1.5                                                         $M_{\Sun}$\\
$T_{\text{eff}}$    & < 6500 K           & --- & --- & < 6250 K\\
$v\sin i$    & ---                          & --- & --- & < 5.0 km/s\\
$\log R'_{\text{HK}}$ & ---                          & --- & --- & < -4.7\\
\enddata
\tablecomments{Filters applied to 2045 TESS systems to produce the Distant Giants sample. TKS filters are taken from \cite{Chontos2021}. Although other filters were applied to produce the TKS sample, we show only those used in our survey's target selection process. MS and SG refer to main sequence and subgiant stars, respectively.}
\end{deluxetable*}
\begin{deluxetable*}{ccrrrrr}
\tabletypesize{\footnotesize}
\tablecolumns{8}
\tablewidth{0pt}
\tablecaption{Distant Giants Sample}

\label{tab:dg_transit_table}

\tablehead{
    \colhead{TOI} &
    \colhead{CPS Name} &
    \colhead{$V$} &
    \colhead{RA (deg)} &
    \colhead{Dec (deg)} &
    \colhead{$R_p \, (\rearth)$} &
    \colhead{P (days)}
}
\startdata
465 &      WASP156 &  11.6 &  32.8 &  2.4 & 5.6 &     3.8 \\
 509 &        63935 &   8.6 & 117.9 &  9.4 & 3.1 &     9.1 \\
1173 &      T001173 &  11.0 & 197.7 & 70.8 & 9.2 &     7.1 \\
1174 &      T001174 &  11.0 & 209.2 & 68.6 & 2.3 &     9.0 \\
1180 &      T001180 &  11.0 & 214.6 & 82.2 & 2.9 &     9.7 \\
1194 &      T001194 &  11.3 & 167.8 & 70.0 & 8.9 &     2.3 \\
1244 &      T001244 &  11.4 & 256.3 & 69.5 & 2.4 &     6.4 \\
1246 &      T001246 &  11.6 & 251.1 & 70.4 & 3.3 &    18.7 \\
1247 &       135694 &   9.1 & 227.9 & 71.8 & 2.8 &    15.9 \\
1248 &      T001248 &  11.8 & 259.0 & 63.1 & 6.6 &     4.4 \\
1249 &      T001249 &  11.1 & 200.6 & 66.3 & 3.1 &    13.1 \\
1255 &     HIP97166 &   9.9 & 296.2 & 74.1 & 2.7 &    10.3 \\
1269 &      T001269 &  11.6 & 249.7 & 64.6 & 2.4 &     4.3 \\
1272 &      T001272 &  11.9 & 199.2 & 49.9 & 4.3 &     3.3 \\
1279 &      T001279 &  10.7 & 185.1 & 56.2 & 2.6 &     9.6 \\
1288 &      T001288 &  10.4 & 313.2 & 65.6 & 4.7 &     2.7 \\
1339 &       191939 &   9.0 & 302.0 & 66.9 & 3.2 &     8.9 \\
1410 &      T001410 &  11.1 & 334.9 & 42.6 & 2.9 &     1.2 \\
1411 &      GJ9522A &  10.5 & 232.9 & 47.1 & 1.4 &     1.5 \\
1422 &      T001422 &  10.6 & 354.2 & 39.6 & 3.1 &    13.0 \\
1437 &       154840 &   9.2 & 256.1 & 56.8 & 2.4 &    18.8 \\
1438 &      T001438 &  11.0 & 280.9 & 74.9 & 2.8 &     5.1 \\
1443 &      T001443 &  10.7 & 297.4 & 76.1 & 2.1 &    23.5 \\
1444 &      T001444 &  10.9 & 305.5 & 70.9 & 1.3 &     0.5 \\
1451 &      T001451 &   9.6 & 186.5 & 61.3 & 2.5 &    16.5 \\
1469 &       219134 &   5.6 & 348.3 & 57.2 & 1.2 &     3.1 \\
1471 &        12572 &   9.2 &  30.9 & 21.3 & 4.3 &    20.8 \\
1472 &      T001472 &  11.3 &  14.1 & 48.6 & 4.3 &     6.4 \\
1611 &       207897 &   8.4 & 325.2 & 84.3 & 2.7 &    16.2 \\
1669 &      T001669 &  10.2 &  46.0 & 83.6 & 2.2 &     2.7 \\
1691 &      T001691 &  10.1 & 272.4 & 86.9 & 3.8 &    16.7 \\
1694 &      T001694 &  11.4 &  97.7 & 66.4 & 5.5 &     3.8 \\
1710 &      T001710 &   9.5 &  94.3 & 76.2 & 5.4 &    24.3 \\
1716 &       237566 &   9.4 & 105.1 & 56.8 & 2.7 &     8.1 \\
1723 &      T001723 &   9.7 & 116.8 & 68.5 & 3.2 &    13.7 \\
1742 &       156141 &   8.9 & 257.3 & 71.9 & 2.2 &    21.3 \\
1751 &       146757 &   9.3 & 243.5 & 63.5 & 2.8 &    37.5 \\
1753 &      T001753 &  11.8 & 252.5 & 61.2 & 3.0 &     5.4 \\
1758 &      T001758 &  10.8 & 354.7 & 75.7 & 3.8 &    20.7 \\
1759 &      T001759 &  11.9 & 326.9 & 62.8 & 3.2 &    37.7 \\
1773 &        75732 &   6.0 & 133.1 & 28.3 & 1.8 &     0.7 \\
1775 &      T001775 &  11.6 & 150.1 & 39.5 & 8.1 &    10.2 \\
1794 &      T001794 &  10.3 & 203.4 & 49.1 & 3.0 &     8.8 \\
1797 &        93963 &   9.2 & 162.8 & 25.6 & 3.2 &     3.6 \\
1823 & TIC142381532 &  10.7 & 196.2 & 63.8 & 8.1 &    38.8 \\
1824 &      T001824 &   9.7 & 197.7 & 61.7 & 2.4 &    22.8 \\
2088 &      T002088 &  11.6 & 261.4 & 75.9 & 3.5 &   124.7 \\
\enddata
\tablecomments{Properties of the 47 stars in the Distant Giants sample, plus the periods and radii of their inner companions. For multi-transiting systems, we checked planets in the order that TESS detected them, and show the properties of the first one that passed the filters in Table \ref{tab:cuts}. Period precisions are truncated for readability. Median uncertainties are as follows: $R_p$---9.6\%; P---60 ppm. Values retrieved from \cite{Chontos2021}.}
\end{deluxetable*}

\begin{figure*}[t]
  \centering\includegraphics[height=1.15\textwidth]{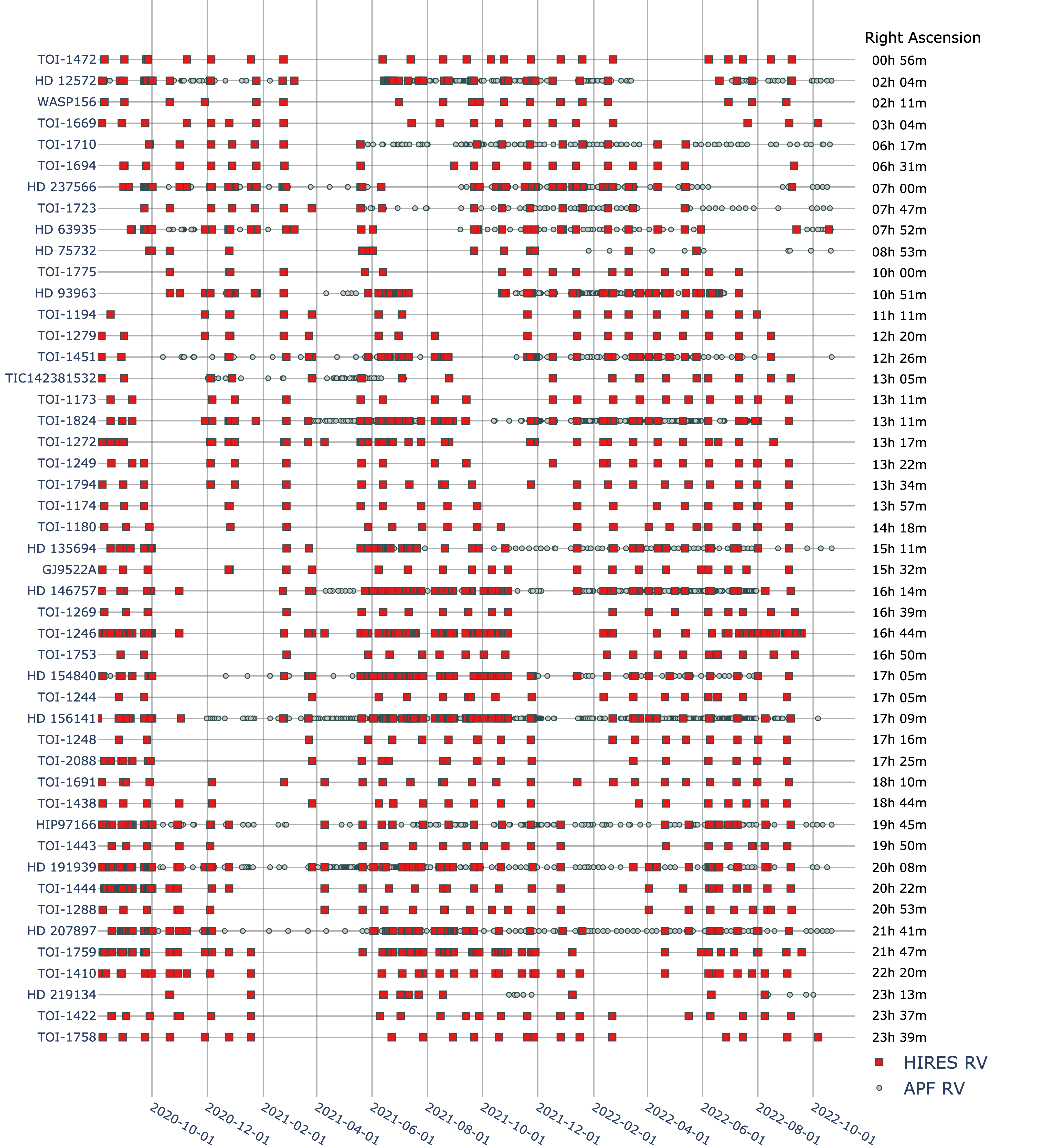}
 \centering\caption{Observations of the Distant Giants sample between its official start in August 2020 and late-2022. Red squares are HIRES RVs and gray circles are APF RVs. Targets are ordered by right ascension, shown in the right margin. The typical target in our sample is inaccessible from Keck Observatory for about three months out of the year, which is reflected in the bands of decreased observation density which run diagonally through the plot. During their observing seasons, all targets generally meet or exceed the prescribed monthly cadence.}
  \label{fig:abacus}
\end{figure*}

\subsection{Observing Strategy}
\label{subsec:observing_strategy}

We tailored the Distant Giants observing strategy to planets with periods $\gtrsim 1$ year and masses above $\sim100 \, \mearth$. Such planets require neither high observing cadence nor high SNR to detect; however, they do require a longer observing baseline than shorter-period planets. In order to maximize Distant Giants's sensitivity to long-period planets within a fixed telescope award, we traded observational cadence and precision for a greater survey duration and target pool. We adopted the following observing strategy: we obtain one observation per target per month. We use the HIRES exposure meter to obtain a minimum SNR of 110 per reduced pixel on blaze at 5500 \AA. The procedure we use to derive radial velocities and uncertainties from raw spectra is described in \cite{ButlerMarcy1996}. The $\sim800$ spectra we have collected for Distant Giants targets at this SNR have a median statistical uncertainty of $1.7$ m/s. Adding our statistical uncertainties in quadrature with the $\sim$2 m/s instrumental noise floor of HIRES \citep{FultonThesis2017}, we estimate a typical RV uncertainty of 3 m/s. We are conducting monthly observations of each target until we have attained 30 observations over 3 years; the six-month surplus compensates for weather losses and target observability seasons. This baseline will allow us to resolve the full orbits of planets with orbital periods $\lesssim 3$ years and sense partial orbits of longer-period companions. Although our sample includes a few legacy RV targets with multi-decade baselines (\citealt{Fischer2014}, \citealt{Rosenthal2021a}), uniform observation of the Distant Giants sample began near mid-2020. The observing baselines of our full sample over a 200-day period are shown in Figure \ref{fig:abacus}.

For targets brighter than $V = 10$, we also obtain observations with the Levy spectrograph on the 2.4-meter Automated Planet Finder telescope (APF, \citealt{Vogt2014}). However, because most of our targets are too dim to benefit from APF observations, we maintain monthly cadence with Keck/HIRES for all targets. This will allow us to analyze our final time series both with and without APF RVs to avoid biasing our planet sensitivity toward brighter stars.

\subsubsection{Existing Observations}

Our target selection process was agnostic to observations collected before the beginning of the survey, which many of our targets possess. After assembling our sample, we examined each target's observing history to determine whether any prior observations could be applied to our survey. We identified three types of systems in our sample:

\begin{enumerate}
    \item \textit{No existing baseline.} 28 targets did not have any useful RV baseline before the beginning of the survey. For our purposes, a useful baseline consisted of observations meeting or exceeding our requirements of monthly cadence and SNR=110, leading up to $\sim$July 2020. We have maintained at least monthly cadence for these targets since Distant Giants began. One of these targets, HD 207897, has RVs as early as 2003, but monthly monitoring only began with our survey.
    
    \item \textit{Partial existing baseline.} 17 targets already possessed a useful RV baseline before the beginning of the survey. These targets will reach their observation quota before those in the subset above.
    
    \item \textit{Finished prior to the survey.} 2 targets, HD 219134 and HD 75732, passed all the cuts in Section \ref{subsec:target_selection} and had already received 30+ observations over 3+ years at the beginning of the survey. We therefore include them in our sample and statistical analysis, but do not obtain further observations. We emphasize that although both HD 219134 and HD 75732 are known to host outer companions, this had no influence on their inclusion in our sample. Had any legacy RV targets exhibiting \textit{non}-detections passed our cuts, they would have been selected as well.
\end{enumerate}

\subsubsection{Surplus Observations and/or Baseline}
Because our survey is carried out under the broader umbrella of TKS, a subset of our targets are observed according to other science objectives with higher cadence requirements. We found that 25 of our selected targets receive more than one observation per month on average, and thus have a greater sensitivity to planets than the remainder of our sample. These systems will require special consideration in our final statistical analysis to correct for their higher planet sensitivity.

In addition to surplus cadence, HD 219134 and HD 75732 have useful RV baselines of nearly 30 years. Their long-period giants, HD 219134 g and HD 75732 d (5.7 years and 14.4 years, respectively), might not have been detectable using the Distant Giants observing strategy. Our prior knowledge of these planets highlights the importance of completeness corrections to account for long-period companions missed due to insufficient observing baseline and/or cadence. Moreover, our detections of HD 219134 g and HD 75732 d will help to characterize completeness in the rest of our systems.

\subsection{RV Observations}
\label{subsec:observations}

We take RV observations according to the standard procedure of the California Planet Search (CPS; \citealt{Howard2010a}). We use the HIRES spectrometer \citep{Vogt1994} coupled to the Keck I Telescope to observe all Distant Giants targets. We place a cell of gaseous iodine in the light path to project a series of fiducial absorption lines onto the stellar spectrum. These references allow us to track the instrumental profile and precisely wavelength-calibrate the observed spectra. For each star, we collected a high SNR iodine-free "template" spectrum. The template, together with the instrumental point-spread function (PSF) and iodine transmission function, is a component of the forward model employed by the CPS Doppler analysis pipeline (\citealt{Howard2010a, ButlerMarcy1996}).

In the first two years of our survey, we resolved the full orbits of giant planets in two of our 47 systems: TOI-1669 and TOI-1694. Although two more systems, HD 219134 and HD 75732, host resolved companions, these planets were detected using hundreds of RVs over multi-decade baselines; further analysis is needed to determine whether they would have been detectable using our observing strategy alone. Finally, seven systems show non-periodic RV trends. We discuss these trends briefly in the conclusion to this paper (\S \ref{sec:conclusion}) and will treat them fully in future work. The parameters of the companions in the four resolved systems are given in Table \ref{tab:resolved_table}, and their stellar parameters are given in Table \ref{tab:dg_stellar_table}.

\subsection{\textit{TESS} Detections of TOI-1669.01 and TOI-1694 b}
\label{subsec:tess_observations}

The SPOC conducted a transit search of Sector 19 on 17 January 2020 with an adaptive, noise-compensating matched filter \citep{Jenkins2002, Jenkins2010, Jenkins2020}, detecting a transit crossing event (TCE) for TOI-1669. An initial limb-darkened transit model was fitted to this signal \citep{Li2019} and a suite of diagnostic tests were conducted to evaluate whether it was planetary in nature \citep{Twicken2018}. The transit signature was also detected in a search of Full Frame Image (FFI) data by the Quick Look Pipeline (QLP) at MIT \citep{Huang2020a, Huang2020b}. The TESS Science Office (TSO) reviewed the vetting information and issued an alert on 29 January 2020 \citep{Guerrero2021}.The signal was repeatedly recovered as additional observations were made in sectors 20, 25, 26, 52, and 53, and the transit signature of TOI-1669.01 passed all the diagnostic tests presented in the DV reports. The source of the transit signal was localized within $4.925 \pm 4.5$ arcseconds of the host star.

The transit signature of TOI-1694 b was identified in a SPOC transit search of Sectors 19 on 17 January 2020. It passed all the DV diagnostic tests and was alerted by the TESS Science Office on 29 January 2020. It was redetected in a SPOC multisector transit search of Sectors 19 and 20 conducted on 5 May 2020, and the difference image centroiding test located the source of the transits to within $0.8 \pm 3.0$ arcsec of the host star.

\begin{deluxetable*}{cccccccc}
\tablewidth{0pt}
\caption{Resolved Distant Giants planet properties}
\label{tab:resolved_table}

\tablehead{
            \colhead{} &
            \colhead{} &
            \colhead{Transiting Planet} \vspace{-0.4cm} &
            \colhead{Transiting Planet} &
            \colhead{Transiting Planet} &
            \colhead{Giant Planet} &
            \colhead{Giant Planet} &
            \colhead{RV Parameter}\\
            \colhead{TOI} \vspace{-0.4cm} &
            \colhead{CPS Name} &
            \colhead{} &
            \colhead{} &
            \colhead{} &
            \colhead{} &
            \colhead{} &
            \colhead{}\\
            \colhead{} &
            \colhead{} &
            \colhead{Period (days)} &
            \colhead{Radius (\rearth)} &
            \colhead{Mass (\mearth)} &
            \colhead{Period (days)} &
            \colhead{Mass (\mj)} &
            \colhead{Reference}}
\startdata
1469 & 219134 & $\pBb$ & $\rBb$ & --- & $\pBc$ & $\mBc$ & 1, 2\\
1669 & T001669 & $\pCb$ & $\rCb$ & $\mCb$ & $\pCc$ & $\mCc$ & 3\\
1694 & T001694 & $\pDb$ & $\rDb$ & $\mDb$ & $\pDc$ & $\mDc$ & 3\\
1773 & 75732 & $\pEb$ & $\rEb$ & --- &$\pEc$ & $\mEc$ & 2\\
\enddata
\tablerefs{(1) \cite{Vogt2015}, (2) \cite{Rosenthal2021a}, (3) This work}
\tablecomments{TOI-1669 and TOI-1694 host newly-discovered distant giant planets. HD 219134 and HD 75732 also host outer giants which were discovered over a much longer baseline. We quote the transiting planet parameters from the TESS DV reports.}
\end{deluxetable*}
\begin{deluxetable*}{ccrcccccc}
\tabletypesize{\footnotesize}
\tablecolumns{8}
\tablewidth{0pt}
\tablecaption{Resolved Distant Giants stellar properties}

\label{tab:dg_stellar_table}

\tablehead{
    \colhead{TOI} &
    \colhead{CPS Name} &
    \colhead{$V$} &
    \colhead{B-V} &
    \colhead{$M_{\star}$ ($M_{\odot}$)} &
    \colhead{$T_{\text{eff}}$ (K)} &
    \colhead{$\log$ g} &
    \colhead{[Fe/H]} &
    \colhead{$v \sin i$ (km/s)}
}

\startdata
1469 &  219134 &  5.57 & 1.02 & $0.79\pm0.03$ & $4839.5\pm100.0$ & $4.48\pm0.10$ & $0.11\pm0.06$ & $0.6\pm1.0$ \\
1669 & T001669 & 10.22 & 0.76 & $1.00\pm0.05$ & $5542.3\pm100.0$ & $4.28\pm0.10$ & $0.26\pm0.06$ & $0.6\pm1.0$ \\
1694 & T001694 & 11.45 & 0.76 & $0.84\pm0.03$ & $5066.4\pm100.0$ & $4.53\pm0.10$ & $0.12\pm0.06$ & $1.2\pm1.0$ \\
1773 &   75732 &  5.95 & 0.86 & $0.97\pm0.05$ & $5363.3\pm100.0$ & $4.31\pm0.10$ & $0.42\pm0.06$ & $0.2\pm1.0$ \\
\enddata
\tablecomments{Stellar parameters for the four stars in our survey hosting fully resolved companions. The effective temperatures, surface gravities, metallicities, and rotational velocities we list here were calculated using \texttt{SpecMatch}, which assigns a fixed uncertainty to each derived parameter. Stellar masses incorporate isochrone constraints, as described in \cite{FulPet2018}.}
\end{deluxetable*}

\section{A Jovian companion to TOI-1669}
\label{sec:TOI-1669}

\subsection{RV Model}
\label{subsec:rv_model_1669}

We visually inspected the time series of the targets in our sample, and found that TOI-1669 exhibited RV variability beyond the noise background. TOI-1669 is a bright ($V = 10.2$) mid-G type solar analog  exhibiting low chromospheric activity ($\log R'_{\text{HK}} = -5.2$) and low rotational velocity ($v \sin i = 0.3$ km/s), as required by our survey filters. Because it is not shared by any other TKS science cases, TOI-1669 was observed according to the Distant Giants observing strategy: we collected $\NobsC$ HIRES spectra between July 2020 and July 2022 at monthly cadence, except during periods when the target was not observable. A subset of TOI-1669's time series is given in Table \ref{tab:time_series}.

We used \texttt{radvel} \citep{radvel} to fit a preliminary model to this system's time series. The model consisted of the inner transiting planet, TOI-1669.01, and the newly-identified outer planet, TOI-1669 b, as well as parameters for linear and quadratic trends and a term characterizing astrophysical and instrumental jitter. We fixed the orbital period (\textit{P}) and time of conjunction (\textit{$T_c$}) of TOI-1669.01 to the values listed in the TESS DV reports. We fit for the three remaining orbital parameters: eccentricity (\textit{e}), argument of periastron (\textit{$\omega$}), and semi-amplitude (\textit{K}).

For TOI-1669 b, we fit for all five orbital parameters with initial values based on visual estimates from the RV time series. We imposed wide priors on free orbital parameters to minimize the bias incurred by our estimates. We used Powell's method \citep{Powell1964} to optimize our likelihood function, and derived parameter uncertainties using Markov Chain Monte Carlo (MCMC) simulations, as implemented in \cite{emcee2013}.

We generated a set of alternative models by excluding different combinations of 1) RV trend/curvature, 2) eccentricity of both planets, and 3) the outer planet itself. We performed a model comparison using the Akaike Information Criterion (AIC; \citealt{Akaike1974}) to find which of these combinations was preferred. The consideration of only one- and two-planet models leaves open the possibility that one or more planets were missed by our analysis. However, the aim of this procedure was not to find every planet in this system, but rather to determine whether it satisfied the basic detection criterion of our survey: hosting at least one outer giant planet in the presence of a close-in small one.

We found that all viable models included both TOI-1669 b and a linear trend. We ruled out models that excluded TOI-1669.01 because this planet was independently confirmed by transit photometry. Eccentricity of either planet improved the model likelihood, but not enough to outweigh the penalty imposed by the AIC for higher model complexity. Quadratic curvature was similarly disfavored. The model preferred by the AIC consists of two planets with circular orbits, as well as a linear trend.

\begin{figure*}
    \centering
    \begin{minipage}{0.485\textwidth}
        \centering
        \includegraphics[width=1.0\textwidth]{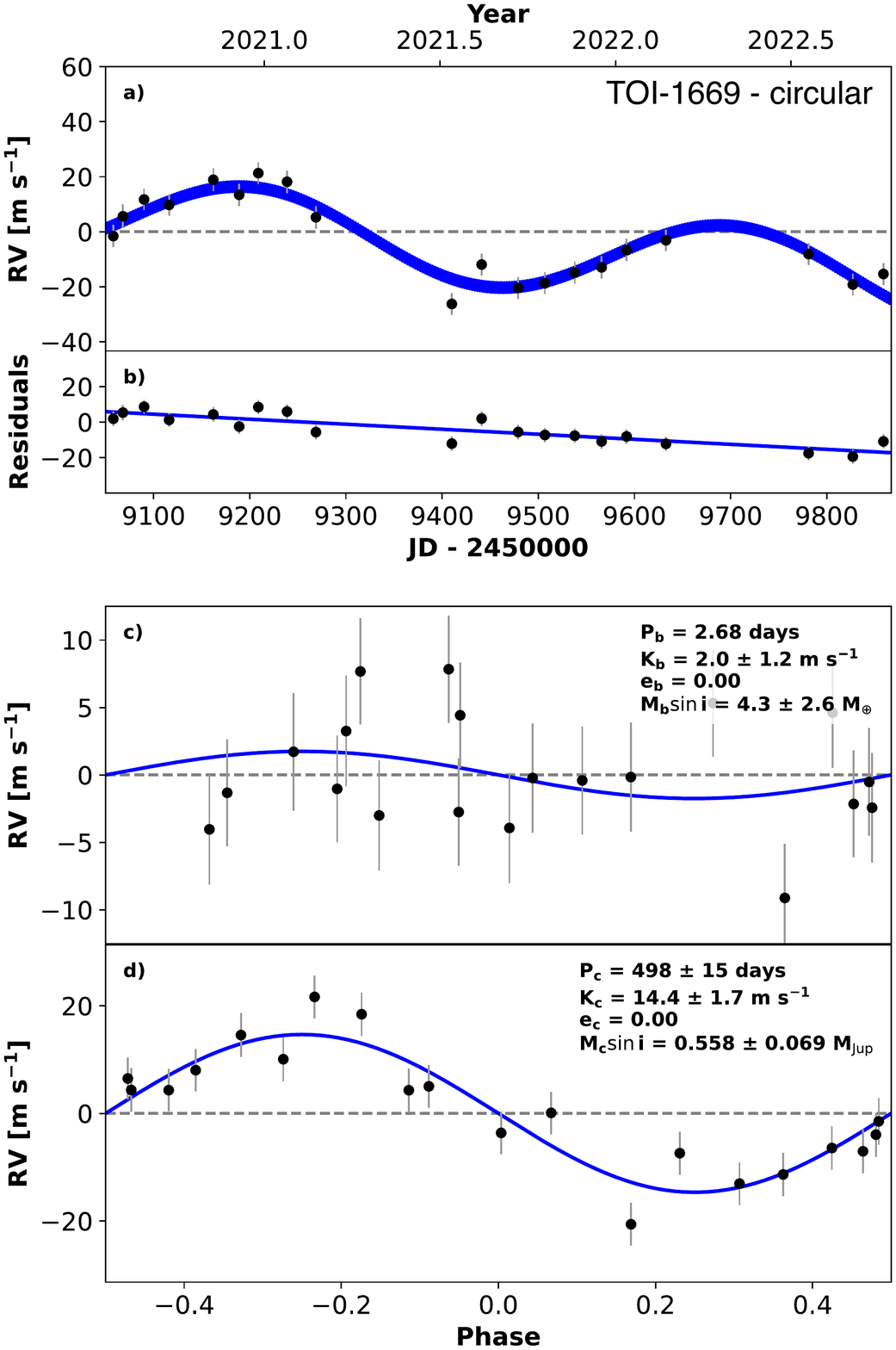} 
    \end{minipage}\hfill
    \begin{minipage}{0.495\textwidth}
        \centering
        \includegraphics[width=1.0\textwidth]{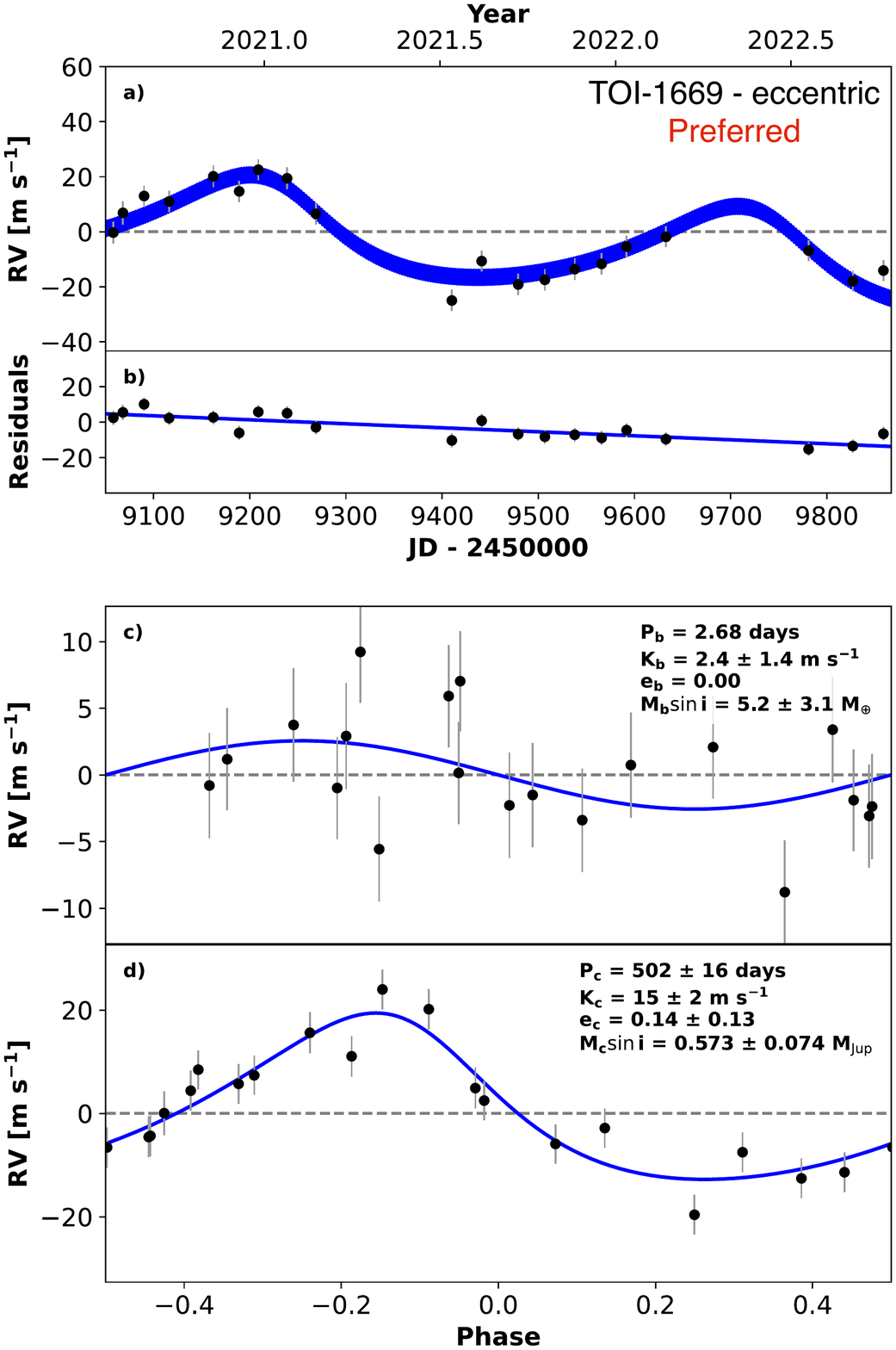} 
    \end{minipage}
    \caption{The RV time series and orbit models for TOI-1669's AIC-preferred circular model (left) and our preferred model including eccentricity (right). Although the circular model has a lower AIC, we adopt the more general model including eccentricity, and emphasize that our limited RV sample contributes to model selection uncertainty. In both figures, \textbf{a)} shows our full Keck/HIRES RV time series (black points) with the fitted model in blue. The residuals to the fit are given in \textbf{b)}. Each subsequent panel shows the time series phase-folded to a particular model planet period. Both models recover consistent periods and masses for the giant TOI-1669 b, as well as a long-term linear trend, suggesting that these parameters are not highly sensitive to our choice of model. By contrast, our eccentric model shows that TOI-1669 b's orbit may deviate from circular. Future observations will resolve this disagreement. The existence of TOI-1669.01 is known from TESS photometry, so we include it in our model despite its low RV amplitude.}
    \label{fig:RV_fits_T001669}
\end{figure*}

We also considered a variation of the AIC-preferred model, with the outer giant eccentricity allowed to vary. Although this model is formally disfavored ($\Delta \text{AIC} = 8.6$), it represents a more realistic scenario than the forced circular case. Moreover, this eccentric model subsumes the circular one, meaning that it could naturally fit a circular orbit for TOI-1669 b if it were favored by the data. The fact that this model instead fits a moderate eccentricity to the outer planet suggests that the AIC-preferred circular fit may not be physical, but rather a result of the AIC's penalization of models with more parameters, which is intended to prevent the over-fitting of small data sets. We adopt the eccentric model and quote its fitted parameters, though we present the circular model alongside it to emphasize the uncertainty in our model selection process. In subsequent sections, we refer to the model with free outer giant eccentricity as "preferred." We show both models together with the full and phase-folded time series in Figure \ref{fig:RV_fits_T001669}.

\begin{deluxetable*}{ccrrcrc}
\tabletypesize{\footnotesize}
\tablecolumns{8}
\tablewidth{0pt}
\tablecaption{Radial Velocities}

\label{tab:time_series}

\tablehead{
    \colhead{TOI} &
    \colhead{CPS Name} &
    \colhead{BJD} &
    \colhead{RV (m/s)} &
    \colhead{RV err (m/s)} &
    \colhead{S-value} &
    \colhead{S-value err}
}

\startdata
1669 & T001669 & 2459537.902 & -13.598 &  1.874 &  0.145 &      0.001 \\
1669 & T001669 & 2459565.861 & -11.912 &  1.919 &  0.146 &      0.001 \\
1669 & T001669 & 2459591.838 &  -5.688 &  1.884 &  0.151 &      0.001 \\
1669 & T001669 & 2459632.799 &  -2.411 &  1.669 &  0.143 &      0.001 \\
1669 & T001669 & 2459781.120 &  -6.998 &  1.671 &  0.136 &      0.001 \\
\hline
1694 & T001694 & 2459591.833 & -13.204 &  1.873 &  0.196 &      0.001 \\
1694 & T001694 & 2459626.823 &  -9.860 &  1.888 &  0.190 &      0.001 \\
1694 & T001694 & 2459654.763 & -40.566 &  1.861 &  0.186 &      0.001 \\
1694 & T001694 & 2459681.789 & -39.828 &  1.663 &  0.201 &      0.001 \\
1694 & T001694 & 2459711.753 & -36.029 &  1.843 &  0.200 &      0.001 \\
\enddata
\tablecomments{We provide subsets of our time series for TOI-1669 and TOI-1694 here for reference. We obtained all RVs for these systems using Keck/HIRES. The full machine-readable versions are available online.}
\end{deluxetable*}

\subsection{False Alarm Probability}
\label{subsec:FAP}

To evaluate the significance of our giant planet detection, we calculated the false alarm probability (FAP) by adapting the procedure of \cite{Howard2010a}. The FAP estimates the probability that a recovered signal arose from random statistical fluctuations rather than an actual planet. We created 1000 "scrambled" versions of TOI-1669's time series by randomly drawing RV values from the original data, with replacement. For each of these data sets, we compared the preferred 2-planet model to the null hypothesis: a model with the inner planet only, with \textit{P} and \textit{$T_c$} fixed and \textit{e}, \textit{$\omega$}, and \textit{K} allowed to vary, and no linear trend. For a given data set, the improvement of the preferred model fit to the data over the single-planet fit to the data is quantified by the difference in $\chi^2$ statistic, $\Delta \chi^2 = \chi^2_{\text{inner}} - \chi^2_{\text{pref}}$, where $\chi^2_{\text{inner}}$ and $\chi^2_{\text{pref}}$ are the minimized $\chi^2$ values of the single-planet and preferred fits to the data, respectively. A more positive $\Delta \chi^2$ value indicates greater performance of the preferred model over the single-planet model. The FAP is simply the fraction of scrambled time series with a greater $\Delta \chi^2$ value than the original time series. Put another way, the FAP gives the fraction of scrambled time series for which the statistical improvement granted by the preferred model over a single-planet model is greater than the improvement granted by the preferred model over a single-planet model for the actual RVs. We found that no scrambled data sets had $\Delta \chi^2$ greater than TOI-1669's original time series, implying an FAP value of less than 0.1\%. We emphasize that this technique quantifies only the probability of false detections as the result of statistical noise, not the probability that one or more planetary signatures were missed in the fitting procedure. However, as we note above, the latter is irrelevant to our search for \textit{at least} one outer giant in each system.

\subsection{Companion Properties}
We recovered a cold sub-Jupiter orbiting TOI-1669 with a period and minimum mass of $\pCc$ days and $M\sin i = \mCc \, \mj$. We found that $e_c = \eCc$, corresponding to a 1$\sigma$ upper limit of 0.27. Our data set for this system is small, consisting of only $\NobsC$ RVs, so we defer precise claims about TOI-1669 b's eccentricity until we have collected more data. Nevertheless, our data is sufficient to indicate that this planet exists and is a distant giant by the standards of our survey.

TOI-1669 also hosts a 2.7-day candidate sub-Neptune detected by TESS. Assuming TOI-1669.01 is a planet, its fitted radius of $R = \rCb \, \rearth$ suggests that it resides on the edge of the Radius Valley, an interval between 1.5 and 2.0 $\rearth$ which exhibits a distinct reduction in planet occurrence \citep{Fulton2017, VanEylen2018}. We derived a mass of $\mCb \, \mearth$ for TOI-1669.01 and used it to calculate a bulk density of $\rhoCb$ g/cc, corresponding to a $1\sigma$ upper bound of $3.19$ g/cc. Due to the uncertainty of our mass measurement, we have not independently confirmed TOI-1669.01 as a planet. Nevertheless, TOI-1669 meets our transit quality requirements (\S \ref{subsec:target_selection}), so we treat TOI-1669.01 as a transiting planet for the purposes of our survey.

Finally, TOI-1669 exhibits a linear trend in the RV residuals. The trends found by our preferred model ($\gdotCecc$ m/s/day) and the circular model ($\gdotC$ m/s/day) are each significant at $> 4\sigma$, and agree with each other to within $1\sigma$. The trend is likely physical, as evidenced by its persistence in both models, and may be caused by an additional long-period companion with a period $\gtrsim2200$ days. To test this hypothesis, we examined direct imaging of TOI-1669 obtained in the I-band (832 nm) with the 'Alopeke speckle imager \citep{Scott2021} coupled to the 8-meter Gemini-North telescope. The imaging reached a roughly constant contrast of $\Delta \text{mag}$ $\approx$ 4 from 0.1"--1.0" and showed no evidence of a luminous companion. This rules out stellar companions $\gtrsim250 \, \mj$ within 100 AU, but leaves open the possibility that a substellar companion orbits TOI-1669 at close separation. If this is the case, extending TOI-1669's observational baseline over the next year will give us greater sensitivity to the companion's orbit.

\section{A Jovian companion to TOI-1694}
\label{sec:TOI-1694}

\subsection{RV Model}
\label{subsec:rv_model_1694}

We also observed significant RV variation in the time series of TOI-1694. TOI-1694 is an early K dwarf  with $V = 11.4$, $\log R'_{\text{HK}} = -5.0$, and $v \sin i = 0.4$ km/s. Like TOI-1669, TOI-1694 was observed at the low cadence prescribed by Distant Giants: we obtained $\NobsD$ HIRES spectra of this target between August 2020 and September 2022. We show a subset of TOI-1694's RV time series in Table \ref{tab:time_series}. 

\begin{figure*}
    \centering
    \begin{minipage}{0.485\textwidth}
        \centering
        \includegraphics[width=1.0\textwidth]{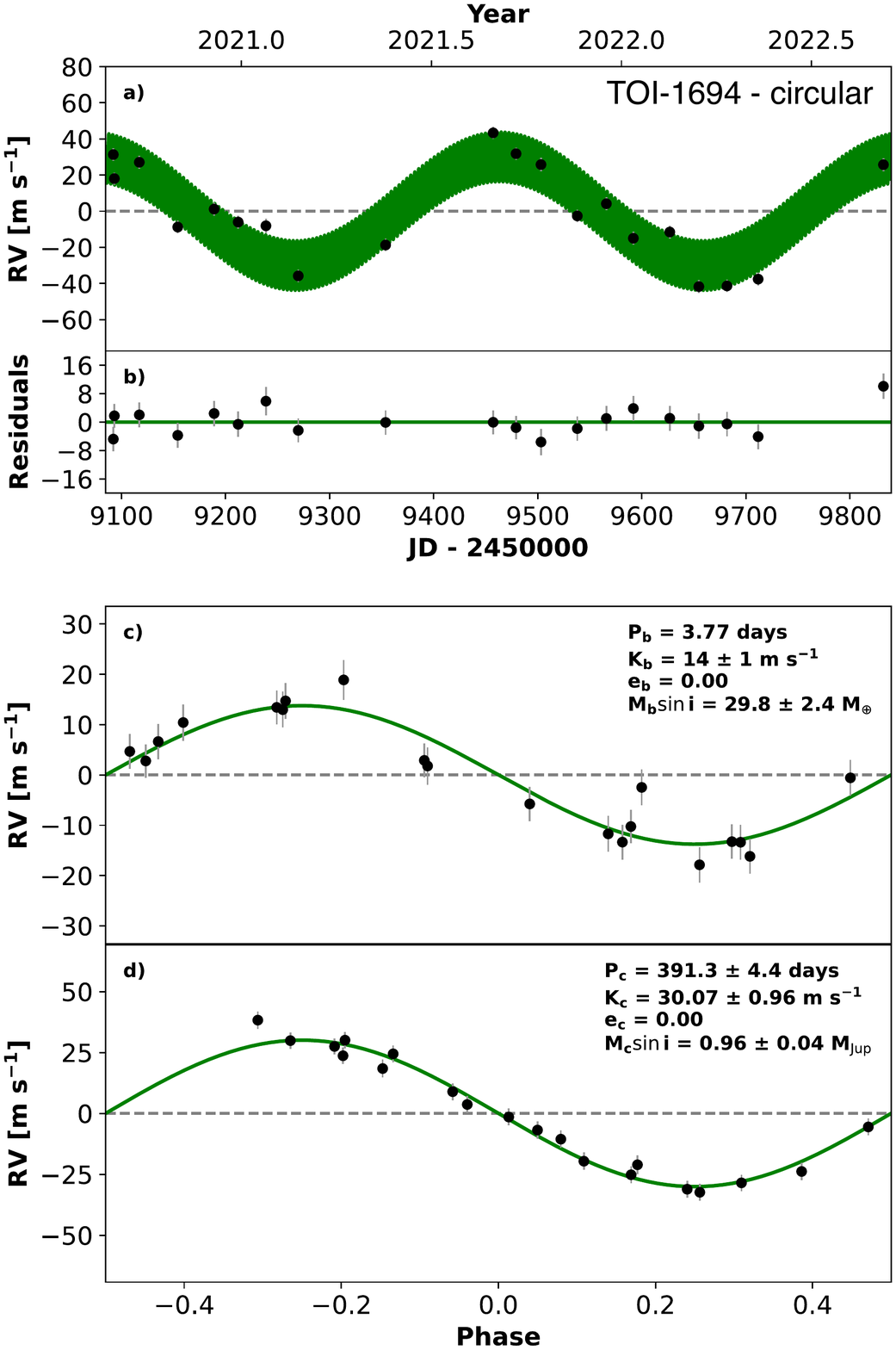} 
    \end{minipage}\hfill
    \begin{minipage}{0.495\textwidth}
        \centering
        \includegraphics[width=1.0\textwidth]{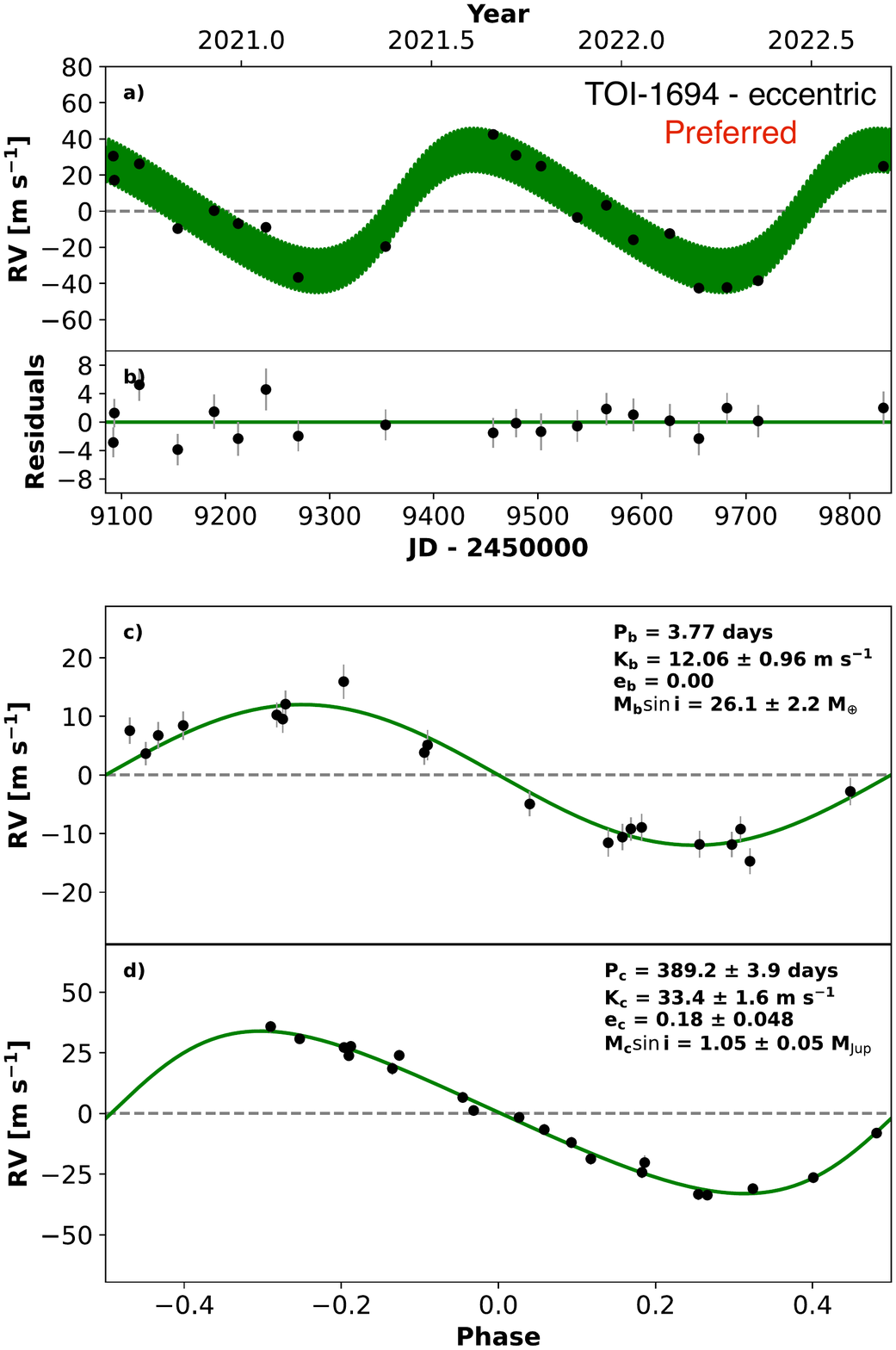} 
    \end{minipage}
    \caption{The RV time series of TOI-1694, together with orbit models assuming a circular (left) or eccentric (right) outer planet. In both figures, \textbf{a)} shows our full Keck/HIRES RV time series (black points) with the fitted model in green. The residuals to the preferred fit are shown in \textbf{b)}. Each subsequent panel shows the time series phase-folded to a particular model planet period. For consistency with our treatment of TOI-1669, we include two models that differ only in the eccentricity of the outer planet. However, in contrast to TOI-1669, the eccentric model we adopted for TOI-1694 is also formally preferred by the AIC. TOI-1694 b's mass and orbital separation identify it as a hot Neptune.}
  \label{fig:RV_fits_T001694}
\end{figure*}

We used the procedure described in Section \ref{subsec:rv_model_1669} to fit an RV model to TOI-1694's time series. We found that a two-planet model with an eccentric outer planet and no trend was preferred over the same model with both orbits forced to circular ($\Delta \text{AIC} = 8.41$). We also tested the preferred model with an added linear trend, and found that the fitted value was consistent with 0 m/s/day, which we interpret as evidence that our model selection process was not heavily influenced by our limited data set. We therefore adopt the AIC-preferred model: an inner planet with a circular orbit, an eccentric outer giant, and no linear trend. Under this model, we calculated FAP$< 0.1\%$ for TOI-1694 c. For consistency with our treatment of TOI-1669, we also present a modified version of the preferred model, with the outer planet's orbit fixed to circular. This model fits similar values to the giant planet's mass and period, suggesting that model uncertainty does not greatly contribute to our overall uncertainty in these parameters. Figure \ref{fig:RV_fits_T001694} shows the preferred RV model for TOI-1694, along with the alternative circular model.

\subsection{Companion Properties}
TOI-1694 hosts an RV-resolved distant giant as well as a TESS-detected inner transiting planet. The outer companion, TOI-1694 c, is a Jupiter analog ($M \sin i = \mDc \, \mj$) with a period of $\pDc$ days and a modest eccentricity of $e = \eDc$.

The inner companion in this system, TOI-1694 b, is a hot super-Neptune with a radius of $\rDb \, \rearth$ and a period of $\sim 3.8$ days. This planet is recovered at high significance by our RV model and has a true mass of $\mDb \, \mearth$. With our mass and radius measurements, we calculate a bulk density of $\rhoDb$ g/cc. TOI-1694 b's low density and large radius suggest that it comprises a rocky core surrounded by a substantial gaseous envelope \citep{Weiss2014, Rogers2015, Fulton2017}. TOI-1694 b is noteworthy because it lies in the Hot Neptune Desert, which refers to the low occurrence of short-period ($P\lesssim 10$ days) planets with masses of $\sim$ 10--100 $\mearth$ \citep{Mazeh2016}.

\begin{figure*}[t]
 \centering\includegraphics[width=1.0\textwidth]{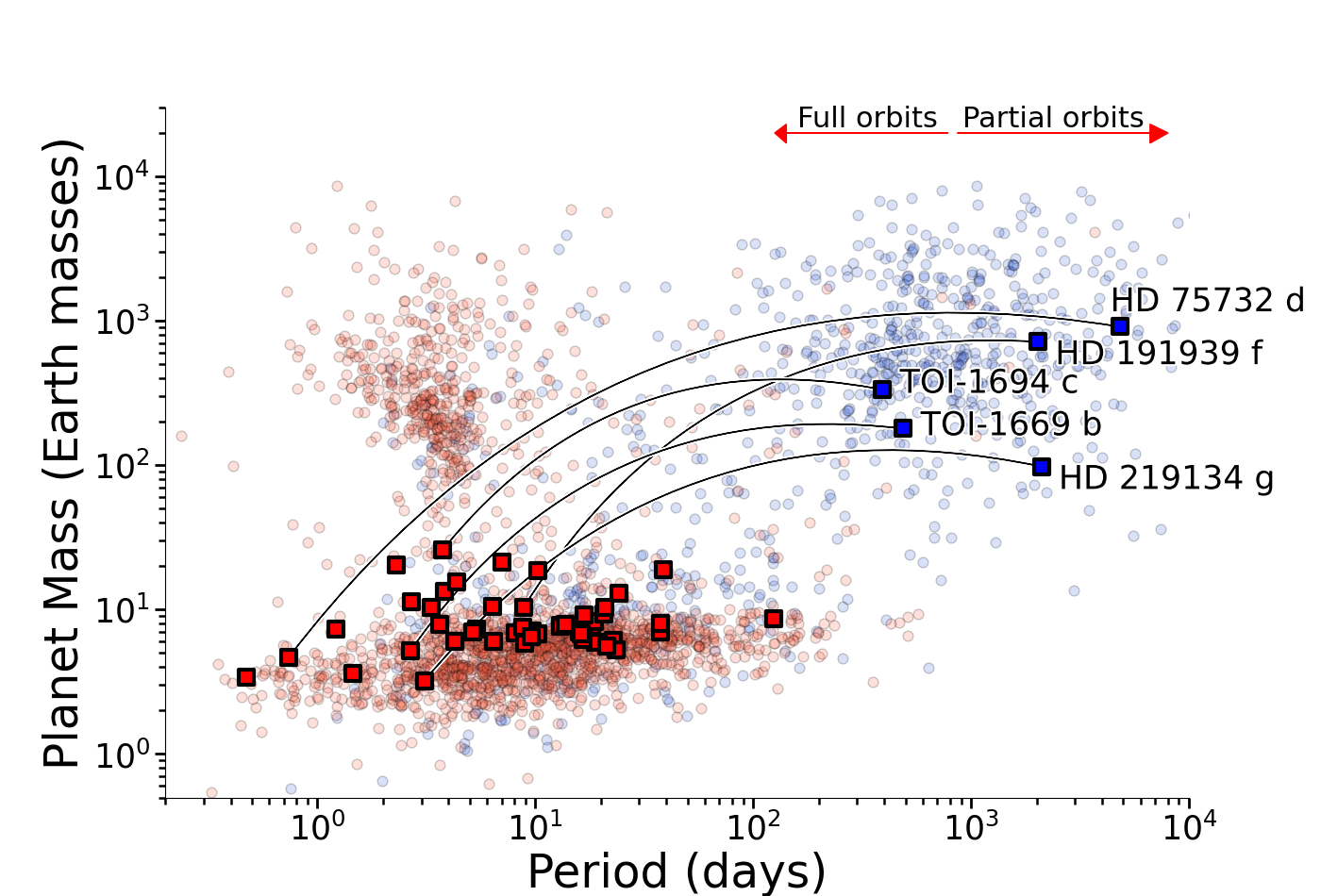}
 \centering\caption{Masses and periods of known exoplanets discovered by RVs (blue circles) and transits (red circles). Transiting planet masses are estimated from their radii using the mass-radius relation of \cite{Weiss2014}. Transiting/resolved Distant Giants companions are overlaid as red/blue squares. Inner and outer companions in the resolved systems, HD 219134, HD 75732, TOI-1669 and TOI-1694, are connected with black lines. We also plot estimated parameters of HD 191939 f, though our analysis of this planet's RV signal is still preliminary. TOI-1669 b and TOI-1694 c have masses and periods typical of cold Jupiters. The red arrows indicate planetary periods shorter and longer than the current $\sim2$-year survey baseline. HD 75732 d and HD 219134 g have orbits well beyond three years, and are only resolved due to their extensive observing histories.}
  \label{fig:mass_per}
\end{figure*}

\begin{figure*}[t]
  \centering\includegraphics[width=1.0\textwidth]{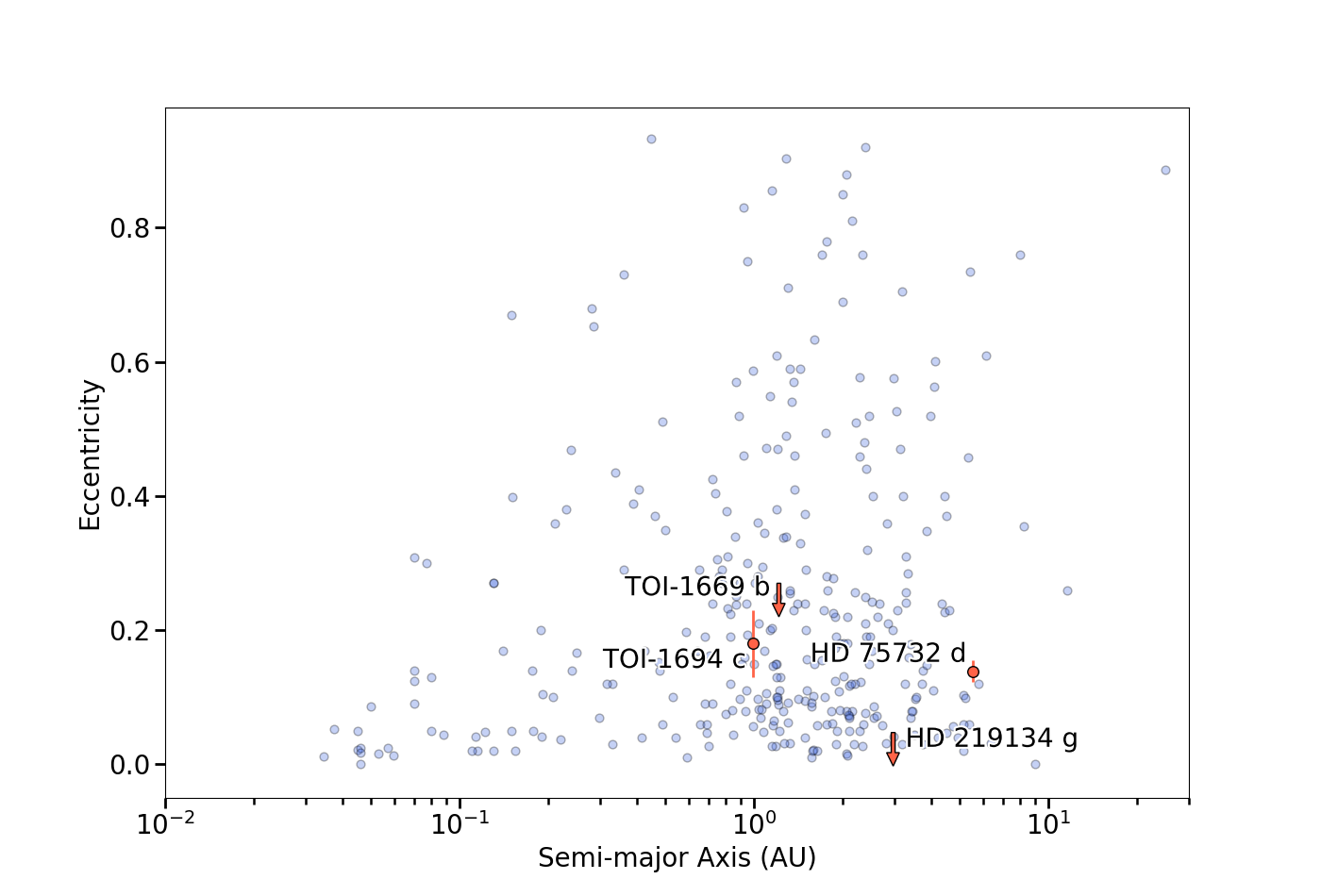}
 \centering\caption{Distribution of eccentricity versus orbital separation for confirmed exoplanets between 0.01 and 30 AU with $\sigma_e \lesssim 0.1$ (blue points). The four resolved giants in our sample are shown in orange. TOI-1669 b and HD 219134 g have eccentricities consistent with zero, and each is represented by an arrow, the base of which shows the $84\%$ upper eccentricity limit. It is not obvious that the distribution from which these planets' eccentricities are drawn is distinct from that of the underlying population.}
  \label{fig:e-a}
\end{figure*}

\section{Discussion}
\label{sec:discussion}

TOI-1669 b and TOI-1694 c are among the first fully resolved outer companions in the Distant Giants sample, along with HD 219134 g \citep{Vogt2015} and HD 75732 d \citep{Fischer2008}, which were known prior to the start of the survey. In a recent analysis of the multi-transiting system HD 191939 \citep{Lubin2021}, we measured a linear trend consistent with a super-Jupiter at a few AU. This detection has become clearer with our extended RV baseline, and we will constrain its parameters more precisely in future work. With our two new giants, two known giants, and a forthcoming characterization of HD 191939's outer giant, we estimate a lower bound of P(DG|CS) $\gtrapprox 5/47$, or roughly $11 \%$, which is comparable to the underlying occurrence rate of $\approx 10\%$ for distant giants out to six-year periods \citep{Cumming2008}. If the six remaining trend systems host distant giant planets and not brown dwarfs or stars, the rate would be $11/47 \approx 23\%$. Because these companions likely have periods much longer than six years, it is more appropriate to compare our $23\%$ estimate to the P(DG)=$17.6^{2.4}_{1.9}\%$ value found by \cite{Rosenthal2021b} for giant planets out to 30-year periods. Using either period limit for distant giants, our preliminary conditional occurrence is similar to the underlying rate. After completing our survey, we will revise our estimate with a full statistical treatment.

In addition to P(DG|CS), the results of our survey will shed light on the period and eccentricity distributions of outer companions to inner small planets. Figure \ref{fig:mass_per} shows the resolved companions in our sample in the mass-period plane. Figure \ref{fig:e-a} compares the resolved companion eccentricities and orbital separations to the underlying population of giant planets. Although we cannot infer population-level traits in either parameter space from these four planets alone, they will serve as a reference in future studies, when we have constrained the properties of more of the companions in our sample.

\section{Conclusions and Future Work}
\label{sec:conclusion}

We presented the Distant Giants Survey, an RV study designed to search for long-period giant companions to inner transiting planets detected by TESS. The objective of Distant Giants is to unify our understanding of two planet classes: the inner small planets discovered in abundance by Kepler, and the Jupiter analogs found by ground-based RV surveys, which are drawn from nearly disjoint stellar samples. In particular, we aim to directly measure P(DG|CS), the conditional occurrence of distant giants in systems hosting a close-in small planet. Our sample consists of 47 inactive Sun-like stars, and our once-per-month observing strategy targets long-period companions. We have completed two years of the three-year survey, allowing us to fully resolve orbits shorter than this baseline.

We reported the discovery of two outer giant planets, TOI-1669 b and TOI-1694 c, identified using Keck-HIRES RVs. We also constrained the masses of each system's inner planet, TOI-1669.01 and TOI-1694 b. TOI-1669 b has a minimum mass consistent with a sub-Jupiter ($M \sin i =  \mCc \, \mj$) with a period of $\pCc$ days. Though our data set is currently too limited to precisely constrain the planet's eccentricity, it is unlikely to be highly eccentric. The inner planet, TOI-1669.01, was recovered at low significance by our RV model, and is probably less than $\sim10 \, \mearth$. TOI-1694 c is a Jupiter analog ($M \sin i = \mDc \, \mj$) with a slightly eccentric orbit ($e = \eDc$) and a period of $\pDc$ days. We recovered the inner planet, TOI-1694 b, at high significance and used the derived true mass of $\mDb \, \mearth$ to calculate a bulk density of $\rhoDb$ g/cc. TOI-1694 b's mass and 3.8-day period place it in the Hot Neptune Desert. Aside from making inroads for dynamical investigation, the coexistence of an inner small planet and an outer giant in these two systems admits them to the subset of unambiguous detections among our sample, which sets the lower bound on our estimate of P(DG|CS).

In addition to TOI-1669 and TOI-1694, long-period giants were already known to orbit HD 219134 and HD 75732 at the beginning of the survey, and we have observed a partial orbit of the outer planet HD 191939 f. We also see linear trends in six RV time series, which we associate with unresolved long-period companions. Combining these groups, we see evidence for distant giants in $\sim23\%$ of our sample. We caution that neither the current number of resolved planets nor the resolved planets plus linear trends should be used for precise calculations of P(DG|CS). There is still a year remaining in the survey, during which new long-period planetary signals could develop and existing trends could be found to be non-planetary. In our final analysis, we will refine the approximation above by computing completeness maps for each target and deriving planet occurrence rates using Poisson point process statistics.

This paper is the first in a series tracking the progress of the Distant Giants Survey. In future work, we will characterize companions discovered during the remaining year of the survey, including HD 191939f, a trend system which we first predicted to be a 6--20-year super-Jupiter through partial orbit analysis \citep{Lubin2021}. The increased phase coverage we have since achieved will let us test our prediction by fitting this object's orbit directly. We will also use this partial orbit analysis to treat the RV trends in six more systems. We will incorporate astrometry and direct imaging to constrain the properties of the objects inducing these accelerations, helping to identify them as planets, brown dwarfs, or stellar companions. Finally, we will refine the relationship between small inner planets and outer giants by using our results to calculate P(DG|CS).

\textit{Facilities:}
Automated Planet Finder (Levy), Keck I (HIRES), Gemini-North ('Alopeke), TESS \\

\textit{Software:}
\texttt{radvel} \citep{radvel}, \texttt{emcee} \citep{emcee2013}, \texttt{SpecMatch} \citep{Petigura2017}.

\section{Acknowledgments}

J.V.Z. acknowledges support from the Future Investigators in NASA Earth and Space Science and Technology (FINESST) grant 80NSSC22K1606. J.M.A.M. is supported by the National Science Foundation Graduate Research Fellowship Program under Grant No. DGE-1842400. J.M.A.M. acknowledges the LSSTC Data Science Fellowship Program, which is funded by LSSTC, NSF Cybertraining Grant No. 1829740, the Brinson Foundation, and the Moore Foundation; his participation in the program has benefited this work. T.F. acknowledges support from the University of California President's Postdoctoral Fellowship Program.

We thank the time assignment committees of the University of California, the California Institute of Technology, NASA, and the University of Hawaii for supporting the TESS-Keck Survey with observing time at Keck Observatory and on the Automated Planet Finder.  We thank NASA for funding associated with our Key Strategic Mission Support project.  We gratefully acknowledge the efforts and dedication of the Keck Observatory staff for support of HIRES and remote observing.  We recognize and acknowledge the cultural role and reverence that the summit of Maunakea has within the indigenous Hawaiian community. We are deeply grateful to have the opportunity to conduct observations from this mountain.  We thank Ken and Gloria Levy, who supported the construction of the Levy Spectrometer on the Automated Planet Finder. We thank the University of California for supporting Lick Observatory and the UCO staff for their dedicated work scheduling and operating the telescopes of Lick Observatory.

Funding for the TESS mission is provided by NASA's Science Mission Directorate. We acknowledge the use of public TESS data from pipelines at the TESS Science Office and at the TESS Science Processing Operations Center at NASA Ames Research Center. This research has made use of the Exoplanet Follow-up Observation Program website, which is operated by the California Institute of Technology, under contract with the National Aeronautics and Space Administration under the Exoplanet Exploration Program.

\clearpage
\bibliography{bib.bib}

\end{document}